
\def\<{\langle} \def\>{\rangle} \def\br{\bf\rm} \def\cl{\centerline}
\def\it{\tenit}   
\def\half{{\scriptstyle{ 1\over 2}}} \def\Im{{\rm Im}} \def\Re{{\rm Re}}
    \def\M{{\br M}}
    
  \def\hs1{\hskip1mm}
\def\h10{\hskip10mm} \def\dag{^\dagger}

\magnification1300

\parskip 2mm plus 1mm \parindent=0pt \def\page{\vfill\eject}

\def\psin{\psi_{\rm in}} \def\psout{\psi_{\rm out}}  \def\alphout{{\alpha_{\rm out}}} \def\H{{\cal H}}


\vskip15mm \bf

\centerline{Quantum State Diffusion: from Foundations to Applications}
{\vskip5mm}
\centerline {by}
\vskip4mm \cl{Nicolas Gisin} \vskip2mm \cl{Group of Applied Physics}
\cl{University of Geneva}
\vskip4mm \cl{and} \vskip4mm \centerline {Ian C Percival} \vskip2mm \centerline
{Department of Physics} \centerline {Queen Mary and Westfield College,
University
of London}

\vskip15mm {Abstract}\rm

Deeper insight leads to better practice. We show how the study of the
foundations
of quantum mechanics has led to new pictures of open systems and to a method of
computation which is practical and can be used where others cannot. We
illustrate
the power of the new method by a series of pictures that show the emergence of
classical features in a quantum world. We compare the development of quantum
mechanics and of the theory of (biological) evolution.

\rm

\vfill

96 January 9th, QMW TH 96- \hfill To celebrate Abner Shimony \eject

\vskip10mm

{\bf 1. Introduction}


Stochastic modifications of the Schr\"odinger equation have been proven to be
useful in a variety of contexts, ranging from fundamental new theories to
practical computations. In this contribution dedicated to Abner Shimony we
recall
the motivations which led us to work in this field and present some thoughts
about what such modifications could mean for future theories. As we shall see,
both of these topics are at the border between physics and philosophy, what
Abner
likes to call "experimental methaphysics".

For Albert Einstein, John Bell and Abner Shimony a quantum theory which can be
used to analyse experiments, but which does not present a consistent and
complete
representation of the world in which we live, is not enough. In recent
years their
ranks have been joined by many others, including those who have proposed
alternative quantum theories to remedy this defect. One such theory is quantum
state diffusion (QSD) [1,2,3,4], which has led to new ways of looking
at laboratory experiments, and correspondingly new {\it numerical} solutions for
open quantum systems. The philosophy of physics has led to useful practical
physics. That such barely physical motivation has led to new concepts and
practical tools is refreshing for physicist like Abner Shimony (and
probably most
of the readers of this book)!

In the next two sections we present our understanding of the quantum measurement
and quantum nonlocality problems and their connections to the QSD model that
developed from these considerations. Then, section 4 presents the QSD model and
section 5 illustrates this model with several examples. This is followed in
section 6 by an illustration of the classical limit of the dynamics of an open
quantum system through the emergence of a strange attractor out of
quantum indeterminism. Finally, in section 7 we introduce an analogy between the
theory of biological evolution and quantum theory, and suggest that the quantum
equivalent of the double helix has still to be found! Actually, in order to take
into account the different views of the authors about the nature of
probabilities,
this section is divided in two subsections entitled "God does not play dice
(NG)"
and "We can't know whether God plays dice (ICP)", respectively. Hence the
title of
this contribution could have been "From Foundations to Applications and back to
Foundations"!

{\bf 2. The quantum measurement problem}


For realists like Abner Shimony and us there is a {\it Quantum measurement
problem.}

A simple quantum measurement results in a single event, usually labeled by
a real
number $\alphout \in R$. It starts with one quantum system in an initial state
$\psin$ in its Hilbert space $\H$. The measurement consists of the emergence of
$\alphout$ under the influence of the quantum system, and since it is a quantum
measurement, a corresponding change in $\psi$ is usually unavoidable. In ideal
cases, the output quantum state of the system for a given value of $\alphout$ is
given by [5]:

$$ \psout = {P_\alphout\psin\over|P_\alphout\psin|}, \eqno(2.1)$$

where $P_\alphout$ is a projection operator.  The probability of this single
event happening is given by the weight:

$$ {\rm Probability}(\alpha) = {\rm Weight}(\alpha) = |P_\alpha\psin|^2.
\eqno(2.2)$$

In all cases in the real world there is one single event with one single result,
that is, one single output. {\it But\hs1} the standard description implies many
outputs, each with its own weight. Next, these weights are identified with the
probability of single events or with the "weights" of world components.

This is the first aspect of quantum measurement. A realist seeks a dynamics
which describes the single event and is consistent with these probabilities. In
the terminology familiar to Abner, one seeks a dynamics which describes (how and
when) the actualization of potential properties.

{\it Nonlinear stochastic dynamics.}

Since (2.1) is nonlinear and since (2.2) is a probability, the dynamics
should be
nonlinear and stochastic. Since the Schr\"odinger equation is linear and
determinate, it cannot define the state of a quantum system in such a realist
theory. Additional terms are needed. This idea appeared in [6], see
also [7]. Abner did also contribute to this program by listing some
desiderata for modifications of the Schrodinger equation [8]

The individual results and their probabilities appear as a result of a dynamical
process, but these conditions are not enough to define the process. For that we
have to consider the strange and apparently contradictory locality properties of
quantum systems. A first aspect of this is the fact that quantum nonlocality can
not be used to send classical information faster than light. This is the
"peaceful coexistence" between quantum nonlocality and relativity, using again
Abner's illuminating terminology [9]. This leads to a strong constraint on
the nonlinear stochastic dynamics, that respects this peaceful coexistence: the
mean quantum state should follow a closed evolution equation. If not, then
"parameter independence" [10,11] would be violated. This constraint was
first met in [12] and Abner was among the very first (and few) who
realized this and encouraged this line of research.

\eject
{\bf 3. Quantum nonlocality and localization}.


The theoretical paper of Einstein, Podolsky and Rosen on elements of reality was
followed by Bell [13] with his inequality for local realistic theories. In
[14] it was shown that the Bell inequality is violated by almost all
quantum states. The papers of Bell led to series of experiments, culminating in
the experiment of Aspect et al [15], which showed that classical events
linked by
a quantum system have nonlocal properties that cannot be reproduced in any
purely
local dynamics. This experimental result does not depend on any theory of
quantum
mechanics, whether it be the usual theory, or an alternative stochastic or
deterministic realistic theory. The experiment shows that the physics is
nonlocal, so {\it every} quantum theory must be nonlocal. The difference between
the usual theory and alternative theories such as QSD is that in the latter the
nonlocality is forced to be explicit.

The nonlocality is accompanied by a reduction or {\it localization} of the wave
packet. For example when a spherical photon wave packet (electromagnetic
wave) is
absorbed by a surface, an electron may be emitted from a very small region of
that surface. In a realistic theory like quantum state diffusion the wave packet
is localized by the interaction with the surface to the region occupied by the
emitted electron.

This is the same localization process that leads to measurement, or the
selection
of a particular state $\alphout$ for a dynamical variable from a spectrum of
possible $\alpha$, and that leads to the localization of the centre of mass of
the Moon, a stone or a Brownian particle. Without this localization the
Schr\"odinger equation disperses the wave function into larger and larger
regions
of position space, and through interactions with other systems, into larger and
larger regions of phase space. This is the second aspect of the quantum
measurement.

The phase space localization which we see with our classical eyes [16] is
reproduced in the dynamics of quantum state diffusion, and this
dynamical localization also makes the equations relatively easy to solve in
practice, by a numerical method that simulates the physical processes seen
in the
laboratory and the rest of the world.

In the foundations of quantum theory the  destruction of coherence, which is
necessary to produce localization or reduction and make the transition from
quantum to classical mechanics, must be distinguished from the dissipation of
coherence. As part of a fundamental theory, the destruction cannot depend on any
division between system and environment.

The localization is due to destruction of coherence. Schr\"odinger dynamics
dissipates coherence, as when a small system like a molecule interacts with a
thermal environment and gets entangled with it. This is represented
mathematically by the change in the reduced density operator for the small
system. But it cannot and does not destroy coherence, because the entanglement
between system and environment is retained if a sufficiently large part of the
environment is included as part of the system. And because the Schr\"odinger
dynamics does not destroy coherence, it cannot produce the localization which is
necessary to make the transition from quantum to classical mechanics. In quantum
state diffusion this coherence {\it is} destroyed, thus allowing the
localization
which is seen in measurement and the dynamics of the classical world.

{\bf 4. Nonlinear stochastic dynamics}


It follows that the dynamics of the quantum state should be stochastic and
nonlinear [17]. This can be achieved by means of deterministic
equations
interrupted by stochastic quantum jumps [18,19,20], or
by a continuous diffusion process, that is, quantum state diffusion [1,3,4].

There are now several alternative quantum theories based on quantum state
diffusion. Following earlier pioneering work of Bohm and Bub [21] and Pearle
[6,22], Gisin [12]  introduced a simple example of quantum state diffusion with
real fluctuations that was generalized by Di\'osi [23] and Gisin [17]. The
complex It\^o form of quantum state diffusion described here was introduced
by us
in 1992 and satisfies a condition of unitary invariance [23,24] and symmetry
[25]. The detailed theory and its applications are given in [1,3,4,16]. The
mathematics of quantum state diffusion also appears in connection with the
theory
of continuous quantum measurements in ordinary quantum theory, as shown, for
example, in Barchielli \& Belavkin [26], Carmichael [20] and references therein.

The original form of the GRW theory of Ghirardi, Rimini and Weber was based on
quantum jumps [27], but it has been reformulated as an intrinsic quantum state
diffusion theory by Di\'osi [28] (first publication of a stochastic Schrodinger
equation related to the GRW model), Pearle [29] (first stochastic Schrodinger
equation for the "raw" ensemble [30]), and Gisin [17] and Ghirardi-Pearle-Rimini
[30] (first stochastic Schrodinger equation for the "physical" ensemble).

The quantum state diffusion theory replaces the deterministic evolution of the
density operator $\rho$ representing an ensemble of open systems,

$$ \dot\rho = -(i/\hbar)[H,\rho]+\sum_j\left(L_j\rho L_j^{\dagger} - \half
L_j^{\dagger}L_j\rho- \half\rho L_j^{\dagger}L_j\right), \eqno(4.1)$$

by a unique stochastic diffusion of a quantum state, representing an individual
system of the ensemble in interaction with its environment. $H$ is the
Hamiltonian, and the $L_j$ are environment operators which represent the
interaction with the environment. The corresponding quantum state diffusion
equation is a stochastic differential equation for the normalised state vector
$|\psi\>$, whose differerential It\^o form is

$$ |d\psi\> = -(i/\hbar)H|\psi\>dt-\half\sum_j(L_j^\dagger L_j +
\ell_j^*\ell_j -
2 \ell_j^* L_j)|\psi\>dt  +\sum_j(L_j-\ell_j)|\psi\> d\xi_j, \eqno(4.2)$$ where
the $\ell_j$ are defined by

$$ \ell_j = \<L_j\> = \<\psi|L_j|\psi\>. \eqno(4.3)$$

The stochastic fluctuation or noise of the diffusion process is all contained in
the standard normalised Wiener fluctuation terms $d\xi_j$, which are of order
$(dt)^\half$ and which satisfy the relations

$$ d\xi_jd\xi_k = 0,\h10 d\xi_j^*d\xi_k = \delta_{jk}dt, \eqno(4.4)$$

$$ \M d\xi_j = 0. \eqno(4.5)$$

The state vector executes a Brownian motion on the unit sphere in Hilbert space.
Each environment operator $L_j$ contributes to the Brownian motion in just two
real directions, and the diffusion is isotropic in this two-dimensional space.
The latter property makes the process (4.2) unique, as illustrated by
examples in
ref. [25,31].

If the density matrix evolution (4.1) and the stochastic pure state evolution
(4.2) have same initial conditions, then for all times the latter defines a
decomposition of $\rho_t$ in terms of a classical mixture of pure states
$\psi_t$:
$$
 \rho_t = M(|\psi_t\>\<\psi_t|),
$$
 where M denotes the mean over the
classical noises $\xi_j$. This relation is basic for the QSD model. It
guarantees
that all means of quantum expectation values $M(\<A\>_{\psi_t}$
agree with standard quantum mechanics. In particular it guarantees that QSD
cannot be used to signal faster than ordinary quantum mechanics, ie not faster
than light. Furthermore, the distribution of pure states determined by (4.2) may
provide deeper insight into the physics of the open system; for instance it
allows us to distinguish dissipation of decoherence from descruction of
coherence.

As a practical method of computation, QSD gains over the direct solution of the
master equation, because for a basis of $N$ states, QSD needs a computer store
with $N$ elements, and the time of computation is also proportional to $N$. For
the direct solution these are proportional to $N^2$. However practical QSD
is run
as a Monte Carlo method, and the computation time increases as the size of the
sample. We have found, however, that a lot can often be learned from a
single run!

The localization by the environment has been demonstrated by many theorems
[3] and numerical examples. For the measurement of a dynamical variable, the
$L_j$ is the operator corresponding to that dynamical variable, and the
individual states localize on one of the eigenspaces of that dynamical variable.
Dissipation and thermal interactions are represented by nonselfadjoint
operators,
and the states tend to localize to wave packets, which are localized in phase
space. Localization in phase space is typical, whereas  localization to an
eigenspace (reduction) is very special.

The phase space localization can be analysed in detail near the semiclassical
limit, where a typical state localizes in three stages, corresponds to three
levels of dispersion in phase space [16]. In the first stage the
density is so much dispersed that it is not confined to a region in which the
dynamical variables can be approximated by dynamically linear variables, that is
by linear combinations of the canonical coordinates and momenta. In the second
stage it is so confined, but the region is large compared to $\hbar^m$ so the
dynamically linear theory applies in its classical version. In the third stage
the localization has effectively confined the system to a region of phase space
comparable to $\hbar^m$, near to the limit imposed by Heisenberg indeterminacy.
It is important to notice that in the early stages there is usually simultaneous
localization in conjugate dynamical variables. There is nothing wrong with this,
since it does not take place near the Heisenberg limit.

Whereas interaction with the environment always tends to localize, the
Schr\"odinger evolution tends to disperse or delocalize the wave function, as in
the example of a free particle wave packet, in which there is a dispersion in
position, and even more strongly in collisions, for which there are outgoing
spherical waves. For nonlinear systems like molecules this dispersion leads to
complicated wave functions that are difficult to compute. In the next section we
demonstrate a computing method which takes advantage of this localization.

So for real open systems there is generally a competition between the localizing
state diffusion and the dispersing Schr\"odinger evolution. Fortunately the
localization is a rapid process, and it often overcomes the Schr\"odinger
dispersion, particularly for large systems, so there is a classical world for us
to live in!

\eject
{\bf 5. Examples and practical applications}

Now we give some examples, mostly from the original QSD papers. Figure 1
shows an
interaction with a measuring apparatus. The full representation would take into
account the detailed physics of this interaction, but just as the important
features of a resistor can be represented by a single real variable, its
resistance [4], so the important features of a quantum measurement can be
represented by a single environment operator, which is proportional to the
dynamical variable being measured. Notice from the QSD equation that for a
environment operator $L_j=L$, the fluctuation term and the corresponding drift
term containing $L$ are both zero when the state is in an eigenstate of $L$.

In Figure 1 it is the energy or photon number of a quantized electromagnetic
field that is being measured, represented by $L=a\dag a$, and the field
starts in
a linear combination of the lowest five odd states. The mean energy is plotted
against time for 9 different runs, which can be thought of as 9 different
experiments: measurements of the system in the same initial state. The energy
starts by fluctuating wildly, and then settles down to one of the energy
eigenstates, with a probability given by the usual quantum formulae of
Section 1.
As we see, the probability of the usual `interpretation' of quantum mechanics,
becomes a probability derived from the dynamics. This equality follows from the
fact that the ensemble of pure states gives the same probabilities as the usual
density operator.

Figure 2 shows an example from quantum optics. It represents a damped oscillator
with an applied sinusoidal resonant force, in interaction representation.
In this
representation the hamiltonian and the single non-selfadjoint annihilation
environment operator are

$$ H=2i(a\dag - a),\h10 L= a, \eqno(5.1)$$

and the system starts in the state $n=8$. The At first the damping and the
fluctuations dominate, but then the system settles down to a state with
negligible oscillation and constant mean energy. This state is a coherent state,
which is a moving Gaussian wave pact. This is an eigenstate of the annihilation
operator, so the fluctuation term is zero, which is why it does not fluctuate.
This is a good example of the localization in phase space which, according to
QSD, gives us our classical world.

Figure 3 illustrates an oscillator in a thermal heat bath [32] in units
for which $\hbar=1$. The four graphs represent the position and momentum, that
continue to fluctuate, as one would expect, and they also illustrate the
standard
deviations as a function of time, and these reduce towards $1/2$, giving the
smallest product permitted by Heisenberg indeterminacy.

Figure 4 shows a single run for a double well potential in a thermal bath
at zero
temperature. This is an example of a `superselection rule' such as occurs for
symmetric molecules. In QSD the system settles down into one well or the other,
as observed by experiment.

One of the surprising things about QSD is that it can represent physical
situations in which there are `quantum jumps'. In QSD they do not happen
instantaneously, but the process is very fast. This is illustrated in Figure 5.

$$ H=0,\h10 L_1 = 6a\dag a,\h10 L_2 = 0.1 a \eqno(5.2).$$

There is simultaneously a damping process represented by the annihilation
operator, and a relatively strong interaction corresponding to a measurement
process represented by $L_1$. Because of the damping process the mean over the
ensemble of the energy or photon number decreases exponentially, but because of
the measurement, the state tends to `try to' stay in the neighborhood of a
particular eigenstate of the energy. The result is that for each run, there
is an
almost constant energy for extended intervals of time, interrupted by a sequence
of jumps. However, the simulation of this process is numerical very inefficient,
so as a practical method QSD does not work well in simulating jumps.

QSD owes a lot of its success as a practical method because of localization, and
is at its best when the localization is strong. This is because the localization
confines the state vectors so that the variance of dynamical variables becomes
smaller, and, in effect, the state vector is confined to a smaller region of
phase space. The quantum state can then be represented in a moving basis (MQSD)
[33,34], which follows this region of phase
space. In practical problems of optics, like second harmonic generation,
this can
save many orders of magnitude in space and time. This gain is over and above the
factor of $N$ gained by representing a state vector instead of a density
operator, as mentioned in the previous section.

This practical success is a direct consequence of the very property of
localization that was introduced in the first place to represent quantum
measurement and classical dynamics in quantum state diffusion as a theory
for the
{\it foundations} of quantum theory. So the study of the foundations of quantum
mechanics has led to new pictures of open systems and to a method of computation
which is practical and can be used where others cannot.

\eject
{\bf 6. Emergence of a classical strange attractor out of a quantum fog.}

In this section we apply the quantum state diffusion model to an open quantum
system whose classical counterpart is chaotic. This provides a nice
illustration of how QSD describes with equations and with figures the
appearance of classical features in a quantum theory.
The system, first introduced in this context by
Spiller and Ralph [35], is a damped, driven,
non-linear oscillator with Hamiltonian (in the
interaction picture, $\hbar=1$):
$$
H=\half\chi a^{\dagger 2}a^2 + iF(t)(a^\dagger -a)
$$
and one environment operator $L=\sqrt{\gamma}a$.
The coefficient $\chi$ represents the anharmonicity and $\gamma$ the friction.
The function $F(t)$ is a periodic string of rectangular pulses defined as
$F(t)=0$ if $t$ mod $\tau < \tau_1$ and $F(t)=F_0$ if $t$ mod
$\tau > \tau_1$,
where $\tau =\tau1+\tau2$ is the period.
We shall use the following values:
$\chi=0.004$, $F_0=2$, $\gamma=0.1$, $\tau_1=5$ and $\tau_2=4.9$.

The corresponding classical dynamical equation reads
$$
{{d\xi}\over{dt}}=-\half\gamma\xi +F(t)-i\chi\xi^2\xi^\ast
\eqno(6.1)$$
where $\xi$ is a complex number whose real and imaginary part represent
position and momentum, respectively.
An interesting invariance property of this system under scaling
allows one to enlarge the
portion of phase space explored by the system during evolution.
More specifically,
there is a one-parameter $\beta$-scaling transformation
$\bar{\xi}=\xi/\beta$, $\bar{\gamma}=\beta\gamma$, $\bar{t}=t/\beta$,
$\bar{F_0}=F_0$,
$\bar{\chi}=\beta^3\chi$ that
does not change the classical equation (6.1), except for a global scaling
of the coordinates $\xi$.
This is relevant for our purpose since in the quantum case
enlarging the explored phase space ($\beta\rightarrow 0$) corresponds
in a natural way to the classical limit. Indeed, since the localization
produced by QSD can not violate the Heisenberg uncertainty relations, the
dimension of the characteristic dimension of
the anharmonic potential relative to $\hbar$ is crucial.
If the wavepacket is localized on the size of $\hbar$ and
if this is relatively small (compared to the potential), then the
wavepacket
remains localized and follows more or less the classical trajectory. If,
on the contrary, the wavepacket remains relatively extended, the classical
dynamics is smeared and purely quantum dynamical features dominate.

This is illustrated on figure 6 which represents four QSD trajectories for
different values of the scaling parameters $\beta$. These trajectories are
represented in phase space at times integer multiple of the period $\tau$.
$\Re\<a\>$ and $\Im\<a\>$ are proportional to position and momentum,
respectively.
In the upper figure $\beta=1$ and the explored phase space is small with
respect to $\hbar$ (recall $\hbar=1$). Not much structure appears, as in
Wheeler's smoky dragon. In the second and third figures $\beta=2$ and $\beta=5$,
respectively. The size of the relevant phase space is larger compared to
$\hbar$ and to the size of the wavepacket
(the latter is closed to the limit set by
Heisenberg's relations). Already some clear structure appears. This
structure looks familiar to people experienced with chaotic classical
systems (6.1) [36]. But, actually, there is no need to study the
classical
equation (6.1). Simply look at the bottom of figure 6 where
$\beta=10$ and the QSD trajectory is almost identical to the classical one. It
corresponds to a strange attractor, a typical feature for an open classical
system. Clearly this classical feature continuously emerges from the
quantum world when the relative dimension of the wavepacket and the
characteristic dimension of the potential get smaller, as can be seen in
figure 6 from the top to the bottom. Note that the
classical system (6.1) has also a fixed point close to (-5,5) which appears in
the two central figures together with random transitions between this
(classically) regular region and the (classically) chaotic region. For
$\beta=10$ these transitions happen only rarely, in particular non are
displayed in the bottom figure 6.
In reference [37] the full classical limit of the QSD equation applied to
this example is presented.

{\bf 7. Quantum State Diffusion, Probabilities and biological evolution.}

In this section we present some views about QSD, the role of probability
in physics and similarities with biological evolution. Since our views
differ substantially, each of us wrote a separate subsection.

{\it 7.1 God does play dice (by NG)}

Let us assume that Nature is nondeterministic: God plays dice.
First, let us emphasize that this would not be the end of science.
Quite the contrary, it was a fresh start for one of the most important
of today's sciences: biology. This
creative time makes the evolution much more interesting. Moreover,
instead of spreading out to infinity, or remaining in a boring
stationary state, as with the Schrodinger equation,
the system localizes dynamically, as in the QSD model.
How would physical laws look like? in particular the laws describing the
(nondeterministic) evolution of physical systems.
I do not know. But it is likely that the evolution equations would incorporate
random numbers. What is a random number:
one does not really know. Actually
it does not really matter. After all one does not know how to prove that
a program is bugfree, but we use programs to compute the number that
deterministic theories predict and we
compare these numbers to experiments. Similarly we could use any reasonable
random number
generator to compute the number of the nondeterministic theory and compare
the statistical predictions to experiments.
(In [38] I have proposed axioms for propensities
(true randomness) in such a way that they are determined by the set of
definite (actual) properties. In this way randomness can be recognized
(contrary to Kolmogorovian randomness) and - moreover - a significant
part of the quantum mechanical Hilbert space structure appears naturally.)

    But then: when and where does chance happen? and what "causes"
it? And what happens to the correlations that interactions between
systems creates? Let us first consider the last question.

    In classical as well as in quantum mechanics the  correlations
become more and more subtle as time and interactions increase.  In
realistic   situations   one   can   prove   that   quickly    the
correlations
are  so  mixed  up  that  it  is  impossible  to  put  them   into
experimental  evidence.  Hence  one  can  forget  about  them  and
consider only density matrices in quantum mechanics  [39,40]  or
distribution functions  in  phase  space  in  classical  mechanics
[41]. From a pragmatic  point  of  view  one  can  as  well
consider that the correlations  are  not  only  hidden,  but  really
destroyed: there is a  correlation  sink.  The
distinction is particularly sharp in quantum physics.  Either  one
assumes that the correlations (also called  quantum  entanglement)
get only hidden and one is satisfied to prove  that this
assumption can not be falsified. Or one  assumes  the
existence of a correlation sink and investigates the consequences.
The first consequence is clearly  that  the  Schrodinger  equation
would  no  longer  be  the  ultimate  (nonrelativistic)  evolution
equation. This is actually the main argument in favor of the first
alternative, which I phrased on purpose in such a way to underline
that it is not more  scientific  as  the  second  alternative.  In
references [3,42] and [43] arguments  in  favor
of each alternative  have  been  discussed.  In  this article   we
clearly follow the second alternative, mainly because new  physics
is more likely to emerge from new theories than from old  ones!  A
generalization of Schrodinger equation is either deterministic  or
stochastic. Deterministic generalizations, such as Weinberg's [44], have
however been  ruled
out by the requirement  of  keeping  the  "peaceful  coexistence"
between     quantum     mechanics     and     relativity [17,45,46]. This
brings us back to the other questions mentioned above,  concerning
chance.

    Let us now come to the question of  physical  chance  and  its
mathematical description, specially in quantum mechanics. It  seems
that there are only two kinds of possible "causes" to chance:

1. It just  happens, without  any  explanation.  This  requires  a
   "universal random generator" (a God who plays dice) and physical laws
   to exploit this  randomness to shape Nature.

2. It takes place  at  intersections  between  independent  causal
   chains, like in Cournot's thesis [47]. This requires a cut
   somewhere through Nature, in order to guaranty the independence
   of the causal chains, like the quantum/classical cut
   in the Copenhagen interpretation.

Following the first alternative, God plays dice.
How could spontaneous  chance be  described?
The mathematical Wiener process could mimic the universal random  generator:
it is just a stupid Markov process  that  keeps  forgetting  every
thing from its past and condemned to make again and again  similar
random choices! But it  enables  $\psi_t$,  the  state  of  the  physical
system, to acquire a shape and a  localization.  Accordingly,  the
quantum world takes advantage of random chance to evolve into one,
among  many  possible,   classical   looking   state   of   affair
[16], as illustrated in section 6. Notice the
similarity with biological  evolution:  there  the  randomness  is
provided by the accidental (another world  for  random)  mutations
and Nature  takes  advantage  of  these  fluctuations  to  produce
order, and even live. According to Darwinism the random mutations
are independent from
the environment. The latter intervenes only in the selection mechanism.
Similarly in a stochastic version of the Schrodinger equation the
fluctuations $d\xi_t$ could be independent of the environment, the latter
taking advantage of the fluctuation to shape the physical system.

So far in the history of Sciences, people have always looked
for deterministic
theories behind apparently random phenomena. This has been extremely
productive. The idea that God plays dice is that in the future scientists
looking for stochastic theories behind apparently organized phenomena will be
even more productive for Science.

{\it 7.2 We can't tell whether God plays dice (by ICP)}


The theory of probability and its problems are at least
as important in biology as in physics.  In the theory of
evolution, random processes come in universally in the source of
new variation, and also in the mixing of genes in the special case
of sexual reproduction.  We shall be concerned only with the new
variation.

In his `Origin of Species' in 1859 Darwin was unable to specify
the mechanism for the origin of variation in species, and was not
clear as to the causes of variability.  The process of selection,
whether natural or under domestication, acted on whatever
variation there was, to produce new varieties and species.

The mechanism became clearer with Mendel's 1865 rules of
heredity, in terms of factors which we now call genes, which
gradually became accepted after 1900.
The genetic theory was only successfully incorporated
into a general mathematical theory of evolution by natural
selection during the 1940s.  In this theory the new variation is
produced by the random mutation of genes, which must be very
small over a few generations for effective natural selection to
take place.

It was not until it was discovered in 1953 that the DNA double
helix contained the genes, that the physical
basis of genetic variation could be understood.

Compare this with the quantum theories of measurement.

Although the mathematical laws of probability in the usual
Copenhagen theory are clear, the theory is no more clear than
Darwin about the mechanism.  The `shifty split' between the
quantum and classical domains is no better, and no worse, than
Darwin's vagueness about the mechanism of variation in species.
Both the physical and biological theories were powerful and
convincing for the purposes for which they were
introduced, and came to dominate their fields. Both of them were
incomplete.

The stochastic theories of quantum mechanics, like quantum
state diffusion, are analogous to the mathematical theories
of biological evolution of the
1940s. In each case the mechanism is clear, but the cause of the
stochastic fluctuations is not. Just as  genetic mutation must
be slow on the time scale of the generations, and produces fitter
species by natural selection, which cannot be produced by
Mendelian rules alone,  so the process of state diffusion is slow
on the time scales of the Schr\"odinger equation, and gives rise
to classical mechanics by localization, which cannot be produced
by Schr\"odinger dynamics alone.  Just as the process of natural
selection led to the development of the enormous and
beautiful variety of modern species from the simplest
beginnings, so process of state diffusion produces the
classical world from the very different quantum world.

The quantum equivalent of the DNA double helix would be the
experimental detection of the elusive quantum-classical boundary.
Primary state diffusion [48,49] is a
development from QSD that suggests possible experiments to detect
it.  It is much more difficult to find than
the double helix of DNA, but there
have been such enormous advances in experiments on individual
quantum systems, particularly in atom interferometry
[50], that it may just be possible.

What appears to be random at one level of experimental sophistication
may look deterministic at another, and vice versa.
We cannot tell whether God plays dice.

{\it 7.3 Conclusions (by NG and ICP)}

We have shown that is possible for those
who disagree about the philosophy to
work together on the physics! We are both very
happy to dedicate both the physics and the philosophy
to Abner Shimony, who has helped us both.

{\bf Acknowledgments} We thank Todd
Brun, Marco Rigo, R\"udiger Schack and Walter Strunz for helpful
communications and the UK EPSRC and Swiss FNRS for financial support.

\vfill\eject {\bf References}

1. Gisin, N. \& Percival, I. C. 1992a The quantum state diffusion
model  applied to open systems. {\it J. Phys. A} {\bf 25},
5677-5691.

2. Gisin, N. \& Percival, I. C. 1992b Wave-function Approach to
Dissipative Processes: Are there Quantum Jumps? {\it Phys. Lett.
A} {\bf 167},  315-318.

3. Gisin, N. \& Percival, I. C. 1993a Quantum state diffusion,
localization and quantum dispersion entropy. {\it J. Phys. A} {\bf
26}, 2233-2244.

4. Gisin, N. \& Percival, I. C. 1993b The quantum state diffusion
picture of  physical processes.  {\it J. Phys. A} {\bf 26},
2245-2260.

5. Piron, C. 199? {\it M\'ecanique Quantique: Bases et Application}, Presses
        Polytechniques et Universitaires Romandes, 1990.

6. Pearle, P. 1976 Reduction of the state vector by a nonlinear
Schr\"odinger equation. {\it Phys. Rev. D} {\bf 13}, 857-868.

7. Gisin, N. \& Piron, C. 1981 {\it Lett. Math. Phys.} {\bf 5}, 379-385.

8. Shimony, A. 1989  in {\it Philosophical Consequences of
                Quantum Theory}, eds. J.T. Cushing and E. McMullin,
                University of Notre Dame Press, Indiana. See also
        {\it Desiderata for a modified quantum dynamics},
        in {\it Philosophy of Science Association 1990}, 1991.

9. Shimony, A. 1983, in {\it Foundations of Quantum Mechanics in
                the Light of New Technology}, ed. S. Kamefuchi, Phys. Soc.
                Japan, Tokyo, 1983.

10. Jarrett, J.P. 1989  in {\it Philosophical Consequences of
                Quantum Theory}, eds. J.T. Cushing and E. McMullin,
                University of Notre Dame Press, Indiana.

11. Shimony, A. 1984 {\it Controllable and uncontrollable non-locality},
                in Proceedings on {\it Foundations of Quantum Mechanics in
                the light of new technology}, p. 225, ed. Kamefuchi et al.,
                Physical Society of Japan.

12. Gisin, N. 1984 Quantum measurements and stochastic processes.
        {\it Phys.  Rev.  Lett.} {\bf  52}, 1657-1660, see also
        {\bf  53}, 1775-1776.

13. Bell, J.S. 1964 {\it Physics} {\bf 1}, 195.

14. Gisin, N. 1991 {\it Phys. Lett. A} {\bf 154}, 201-202.

15. Aspect, A., Dalibard, J. \& Roger, G. 1982 {\it Phys. Rev. Lett.} {\bf 49},
        1804.

16. Percival, I. C. 1994a Localisation of wide open quantum systems.
{\it J. Phys. A} {\bf 27}, 1003-1020.

17. Gisin, N. 1989 Stochastic quantum dynamics and relativity. {\it
Helv. Phys. Acta} {\bf 62}, 363-371.

18. Diosi, L. 1985 {\it Phys. Lett. A} {\bf 112}, 288;
                {\bf 114}, 451, 1986; {\bf 185}, 5, 1994.

19. Dalibard, J., Castin, Y. \& K. Molmer 1992
       {it Phys. Rev. Letts.}  {\bf 68} 580-583; see also
       J. Opt. Soc. Am. {\bf 10}, 524, 1993.

20. Carmichael, H. J. 1993 {\it An Open systems Approach to Quantum
Optics, Lecture Notes in Physics m18}, Berlin: Springer.

21. Bohm, D.  \& Bub, J. 1966 A proposed solution of the measurement
problem in quantum mechanics by a hidden variable theory.  {\it
Rev. Mod. Phys.}, {\bf 38}, 453-469.

22. Pearle, P. 1979 Towards explaining why events occur. {\it
Internat. J. Theor. Phys.} {\bf 18}, 489-518.

23. Di\'osi, L. 1988a Quantum stochastic processes as models for
state vector reduction. {\it  J. Phys. A} {\bf 21}, 2885-2898.

24. Percival, I. C. 1989 Diffusion of Quantum States 2,
 Preprint QMC DYN 89-4, School of Mathematics, Queen Mary
College London.

25. Gisin, N. \& Cibils, M. 1992 {\it  J. Phys. A}
       {\bf 25}, 5165-5176.

26. Barchielli, A. \& Belavkin, V. P. 1991 Measurements continuous in
time and {\it a posteriori} states in quantum mechanics. {\it J.
Phys. A} {\bf 24}, 1495-1514.

27. Ghirardi, G.-C., Rimini, A. \& Weber, T. 1986 {\it Phys. Rev. D}
{\bf 34}, 470-491.

28. Di\'osi, L. 1988b Continuous quantum measurement and It\^o
formalism. {\it Phys. Lett. A} {\bf 129}, 419-442.

29. Pearle, P. 1989 Combining stochastic dynamical state-vector
reduction with spontaneous localization. {\it Phys. Rev. A} {\bf
39}, 2277-2289.

30. Ghirardi, G.-C., Pearle, P., \& Rimini, A. 1990 {\it Phys. Rev. A}
{\bf 42}, 78.

31. Rigo, M. \& Gisin, N. 1995 Unravelings of the master equation and the
emergence of a classical world {\it Quantum \& Semiclassical Optics}, in press.

32. Spiller T. P.,  Garraway, B. M. \& Percival, I. C. 1993 Thermal
equilibrium in the quantum state diffusion picture
{\it  Phys. Lett. A} {\bf 179}, 63-66.

33. T Steimle, G Alber and I C Percival  1995
Mixed classical-quantal representation for open quantum systems {\it J Phys
A} {\bf 28} L491-496.

34. R Schack, T Brun and I C Percival 1995 Quantum state
diffusion, localization and computation {\it J Phys A}
{\bf 28} 5401-5413

35. Spiller, T. \& Ralph, J.F. 1994 {\it Phys. Lett.} {\bf 194}, 235.

36. Szlachetka, P.  et al. 1993 {\it Phys. Rev. E} {\bf 48}, 101.

37. Gisin, N. \& Rigo, M. 1995 Relevant and Irrelevant Nonlinear Schr\"odinger
        Equations {\it J. Phys. A}, in press.

38. Gisin, N. 1991, Propensities in a nondeterministic Physics, {\it Synthese}
        {\bf 89}, 287-297.

39. Joos, E. \& Zeh, H. D. 1985 The emergence of classical properties
 through interaction with the environment. {\it Z. Phys. B} {\bf
59}, 223-243.

40. Zurek, W.H. 1991 {\it Physics Today}, p36, October.

41. Prigogine, I. \& Stengers, I. 1979 {\it La Nouvelle Aliance}, Gallimard,
Paris.

42. Zeh, H. D. 1993 There are no quantum jumps, nor are there
particles! {\it Phys. Lett. A} {\bf 172}, 189-192.  See also contribution
        to {\it Quantum state diffusion and quantum state localisation},
        eds L. Diosi and B. Lukacs, World Scientific 1993.

43. Gisin, N. \& Percival, I. C. 1993c Stochastic wave equations
versus parallel world components. {\it Phys. Lett. A} {\bf
175},144-145.  See also contribution
        to {\it Quantum state diffusion and quantum state localisation},
        eds L. Diosi and B. Lukacs, World Scientific 1993.

44. Weinberg, S. 1989 {\it Annals Phys.} {\bf 194}, 336-386.

45. Gisin, N. 1990 Weinberg's non-linear quantum mechanics and
supraluminal communications. {\it Phys. Lett. A} {\bf 143}, 1-2.

46. Weinberg, S. 1993 {\it Dreams of a Final Theory},
                Hutchinson Radius, London.

47. Cournot A. 1843 {\it Exposition de la th\'eorie des chances et des
        probabilit's}, Librairie Hachette. Reprinted in part in {\it
        Etudes pour le centenaire de la mort de Cournot}, ed. A. Robinet,
        Edition Economica, 1978.

48. I C Percival 1994b Primary state diffusion {\it Proc. Roy. Soc A}.
{\bf 447} 1-21

49. I. C. Percival 1995 Quantum space-time fluctuations and primary
state diffusion  {\it Proc. Roy. Soc A} {\bf451} 503-513

50. Kasevich, M., \& Chu, S. 1991 Atomic interferometry using
stimulated Raman transitions. {\it Phys. Rev. Letts.} {\bf 67},
181-184.

\page

{\bf Figure captions.}

1. Mean photon number as a function of time for the measurement process.
The stochastic convergence to the eigenstates can be clearly seen.

2. The forced damped linear oscillator, showing reduction towards a coherent state which has
no stochastic fluctuations.

3. Illustration of the approach to thermal
equilibrium of a harmonic oscillator.
The initial state is the pure number state $|3\>$.
The two lines with large oscillations represent the mean position and
momentum, $\< p \>$ and $\< q \>$. The
two lines with smaller oscillations represent
the standard deviations $(\Delta p)^2$ and
$(\Delta q)^2$. This example illustrates how an arbitrary initial state tends
asymptotically to a coherent state. The center of these coherent states
follow then a classical stochastic process [37].

4. Symmetry breaking for a double well (at $x=+8$ and $x=-8$)
potential with two dissipative
environment operators acting independently on each well. The plots for a
single run show the mean position $\< q\>$ and the RMS deviation in
position $\Delta q$. The localization in the $x=+8$ well, the reduction in
the variation and the damping of the stochastic fluctuations are clearly
shown.

5. A quantum cascade with emission only. The continuous state diffusion
automatically produces sudden transitions between quantum states in a
single run, but these are not instantaneous jumps.

6. Emergence of a classical strange attractor out of a quantum fog for a
Kicked, Damped, Anharmonic OScillator (KAOS).
Poincar\'e sections at the period of the driving force are displayed. In the
upper figure the relevant dimension of the potential is large with respect
to $\hbar=1$, hence quantum indeterminacy dominates. In the lower figure,
on the contrary, the potential and damping are scaled such that the
typical dimension in phase space are large with respect to $\hbar=1$, hence
the strange attractor of the classical KAOS is clearly shown.
The two medium figures
correspond to intermediate cases, in which random transitions between
a fixed point and the strange attractor can also be seen.

\end